\documentclass[pdflatex,sn-mathphys]{sn-jnl}% Math and Physical Sciences Reference Style
\normalbaroutside
\usepackage{tikz-cd}
\usepackage{hyperref}

\jyear{2021}%

\theoremstyle{thmstyleone}%
\newtheorem{theorem}{Theorem}%  meant for continuous numbers
\newtheorem{proposition}[theorem]{Proposition}% 

\theoremstyle{thmstyletwo}%
\newtheorem{example}{Example}%

\theoremstyle{thmstylethree}%
\newtheorem{definition}{Definition}%

\begin{document}

\title[The arithmetic topology of genetic alignments]{The arithmetic topology of genetic alignments}


\author[1,2]{\fnm{Christopher} \sur{Barrett}}\email{clb5xe@virginia.edu}
\equalcont{These authors contributed equally to this work.}

\author[1]{\fnm{Andrei} \sur{Bura}}\email{cb8wn@virginia.edu}
\equalcont{These authors contributed equally to this work.}

\author[1]{\fnm{Qijun} \sur{He}}\email{qh4nj@virginia.edu}
\equalcont{These authors contributed equally to this work.}

\author[1]{\fnm{Fenix} \sur{Huang}}\email{fwh3zc@virginia.edu}
\equalcont{These authors contributed equally to this work.}

\author*[1,3]{\fnm{Christian} \sur{Reidys}}\email{cmr3hk@virginia.edu}
\equalcont{These authors contributed equally to this work.}

\affil*[1]{\orgdiv{Biocomplexity Institute}, \orgname{University of Virginia}, \orgaddress{\street{994 Research Park Boulevard}, \city{Charlottesville}, \postcode{22911}, \state{VA}, \country{USA}}}

\affil[2]{\orgdiv{Department of Computer Science}, \orgname{University of Virginia}, \orgaddress{\street{351 McCormick Road}, \city{Charlottesville}, \postcode{22904}, \state{VA}, \country{USA}}}

\affil*[3]{\orgdiv{Department of Mathematics}, \orgname{University of Virginia}, \orgaddress{\street{141 Cabell Drive}, \city{Charlottesville}, \postcode{22904}, \state{VA}, \country{USA}}}

% \affil[2]{\orgdiv{Department}, \orgname{Organization}, \orgaddress{\street{Street}, \city{City}, \postcode{10587}, \state{State}, \country{Country}}}

% \affil[3]{\orgdiv{Department}, \orgname{Organization}, \orgaddress{\street{Street}, \city{City}, \postcode{610101}, \state{State}, \country{Country}}}


\abstract{We propose a novel mathematical paradigm for the study of genetic variation in sequence alignments. 
This framework originates from extending the notion of pairwise relations, upon which current analysis is based on, to 
$k$-ary dissimilarity. This dissimilarity naturally leads to a generalization of simplicial complexes by endowing simplices 
with weights, compatible with the boundary operator. We introduce the notion of $k$-stances and dissimilarity complex, the former encapsulating arithmetic as well as topological structure expressing these $k$-ary relations. We study basic mathematical properties of dissimilarity complexes and show how this approach captures an entirely new layer of biologically relevant viral dynamics in the context of SARS-CoV-2 and H1N1 flu genomic data.}

\keywords{Hamming distance, $k$-stances, sequence dissimilarity, phylogeny, weighted simplicial complexes, weighted algebraic homology}



\maketitle

\section{Introduction}
Genetic variation is the observed difference at the genetic sequence level between individuals in a population and is the key contributor to phenotypic diversity. It affects population dynamics and ultimately the evolution of the entire system.

One of the key tasks of Biology is understanding the development of genetic variation within a given population. Namely, the extraction of evolutionary relationships and histories among the sequences present (i.e their phylogenetics). These relationships are construed in the guise of the phylogenetic tree; a graph topological structure that underpins our understanding of the major evolutionary transitions appearing in the system. This structure is central to inferring everything from the emergence of new body plans, novel metabolism and the origin of new genes to detecting molecular adaptation, understanding morphological character evolution and reconstructing demographic changes in recently diverged species \cite{kapli2020phylogenetic}.

Such a tree is generally constructed from metric (pairwise) information present at the sequence level. For two sequences of equal length, one naive approach is to employ the Hamming distance, which counts the number of positions in the sequence pair with different entries at those positions. For sequences of different length, alignment methods are applied to obtain equal length via the insertion of ``gap'' symbols~\cite{needleman1970general,smith1981identification}. Other metrics can be used, such as the edit distance, which can still be viewed as weighted versions of the Hamming metric~\cite{berger2021levenshtein}.

As a result, metric-based phylogenetic tree constructions integrate only pairwise dissimilarity information. This information can be viewed as a complete graph where each node represents a sequence and the length of an edge represents the dissimilarity between the two corresponding sequences the edge links. The integration process in metric-based phylogeny can be summarized as finding a spanning tree whose inherited metric information ``best fits'' the metric in this complete graph~\cite{felsenstein2004inferring,saitou1987neighbor}.

\begin{figure}[ht]%
\centering
\includegraphics[width=0.5\textwidth]{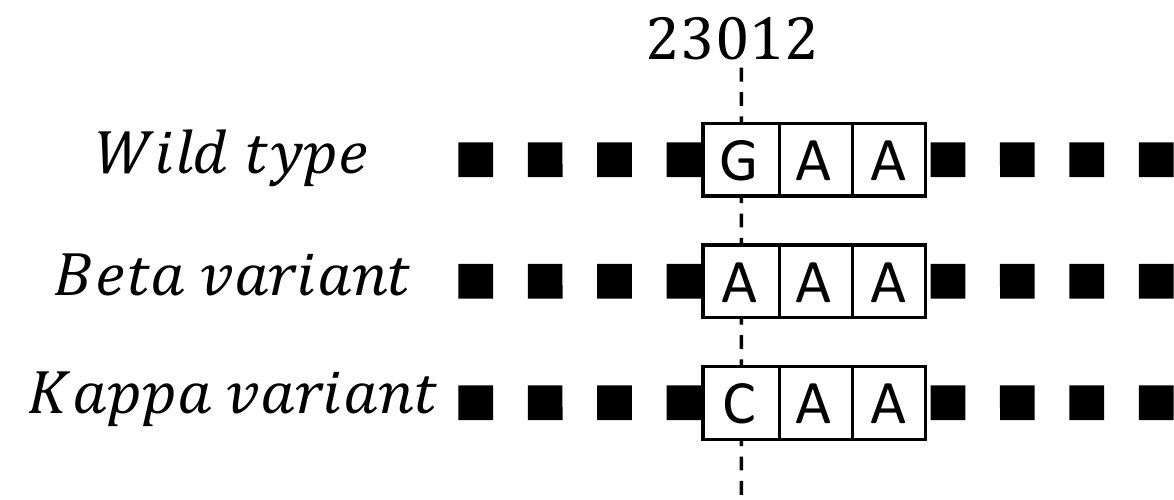}
\caption{Site 23012 on the SARS-CoV-2 genome.}\label{F:covid}
\end{figure}

Graphs are natural choices but can only encode pairwise relations. However in a population of genetic sequences, there exist higher order interactions that cannot be expressed from pairwise relations alone. In a population of sequences, different positions (sites) exhibit different nucleotide diversities. Some sites are conserved allowing almost no polymorphism while others, the key contributors to genetic variation, are less conserved. Polymorphic sites that allow three or more nucleotide realizations are of interest as they are particularly indicative of the population searching the fitness landscape. For instance, in the  SARS-CoV-2 genome, site 23012, located in the spike protein region exhibits such polymorphisms, see Figure~\ref{F:covid}.

\begin{figure}[ht]%
\centering
\includegraphics[width=0.8\textwidth]{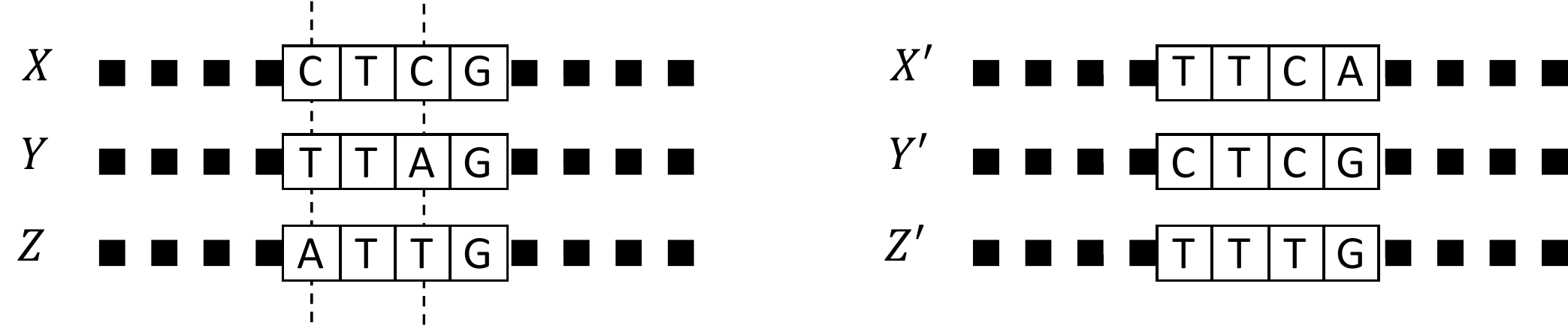}
\caption{Segments of SARS-CoV-2 genomes and their GSAID IDs. X: EPI-ISL-428678, Y: EPI-ISL-509797, Z: EPI-ISL-656231, X': EPI-ISL-467431, Y': EPI-ISL-428678, Z': EPI-ISL-415706. We cannot distinguish the triplet $\{X,Y,Z\}$ from that of $\{X',Y',Z'\}$ if we restrict ourselves to pairwise Hamming distance comparisons. However, $\{X,Y,Z\}$ contains two highly polymorphic sites, while $\{X',Y',Z'\}$ does not contain any.
}\label{F:real}
\end{figure}

This site exhibits the nucleotide G in the wild type sequence, while in the Beta variant of the virus, it mutates to A and 
induces an amino acid change from E to K at position 484. 
E484K is one of the characteristic mutations of the Beta variant and improves the virus's ability to evade the host's immune system~\cite{wise2021covid}. One the other hand, in the Kappa variant, site 23012 mutates to C and results in the amino acid change E484Q, which is thought to enhance ACE2 receptor binding~\cite{cherian2021sars}, and may reduce the vaccine-stimulated antibodies' ability to attach to this altered spike protein~\cite{wilhelm2021antibody}. Key features of these highly polymorphic sites cannot be expressed employing Hamming distance comparisons alone, see Figure~\ref{F:real}.

To capture these multi-fold sequence interactions, we propose a paradigm shift essentially passing from weighted 
graphs to weighted simplicial complexes leading to the concept of \emph{weighted simplicial phylogeny}. 
This is embodied in a novel arithmetico-topological structure, the dissimilarity complex, capable of encoding such sequence hyper-relations. An (unweighted) simplicial complex is classically regarded as just a set composed of points, edges, triangles, tetrahedra and their higher dimensional counterparts. In addition to its combinatorics, a simplicial complex can also be studied from a global (topological) perspective as it naturally gives rise to a topological space. In particular, homological (algebraic) invariants of simplicial complexes often already capture important information about global structural features of the sequence phylogeny. For instance, $k$-dimensional holes in such spaces can be shown to correspond to multi-fold recombination~\cite{chan2013topology}. Recently, weighted homology, an augmentation of simplicial homology that further encodes arithmetical information for each simplex has been developed~\cite{bura2021weighted,dawson1990homology,ren2018weighted}. These weights and the information they carry are central to the new theory as they enrich the algebraic invariants of classical simplicial homology with arithmetic torsion that encodes additional combinatorial information about the sequence phylogeny.

Our paper is organized as follows:

Subsection~\ref{S:sota} provides context by reviewing the current phylogenetic study of a sequence alignment via Hamming distance optimizations and phylogenetic trees.

In Section~\ref{S:kstances} we introduce $k$-stances, a higher dimensional relation between $k$ aligned sequences representing a dissimilarity measurement that naturally generalizes the Hamming distance. Subsection~\ref{S:kstance prop} deals with the mathematical properties of this this $k$-ary relation, both related to metric properties it inherits from the Hamming distance (pairwise case) and to properties that are intrinsic to its higher dimensionality. Subsection~\ref{S:kstance recomb} discusses $k$-stances in the context of genetic recombination.

Section~\ref{S:dissimilarity complex} introduces the dissimilarity complex: an arithmetico-geometrical space whose topological structure encodes the $k$-stance relations among the sequences of an alignment by integration across all $k$-dimensions. We shall briefly review the notion of simplicial complexes and their build principles, usage and means of study. In Subsection~\ref{S:construction of dissimilarity complex} we provide the construction of the dissimilarity complex of a sequence alignment and establish some basic properties. In Subsection~\ref{S:the weighted homology of the dissimilarity complex} we then introduce weighted homology and show how it is connected to the dissimilarity complex of an alignment.

Section~\ref{S:mathematics of single column} analyzes the ``induction basis'' of alignments of length one. We shall show 
here that all $k$-stances are in fact connected.

Section~\ref{S:multi column case} considers the multiple column case. In Subsection~\ref{S:k-stance statistics multi column}
we present two case studies on the statistics of $k$-stances for alignments of SARS-CoV-2 and H1N1 flu genomes, while in Subsection~\ref{S:homology multi column} we hint at the relationships between the arithmetic torsion that arises in the weighted homology of a multi-column alignment and its connections to genetic recombination present in the population 
of the alignment.

Section~\ref{S:discussion} we integrate our results and discuss future directions of work. Finally, Section~\ref{S:proofs} 
contains all proofs.

\subsection{Hamming distance and phylogeny}\label{S:sota}
Let us begin by revisiting the underlying ideas of Hamming distance, its basic metric properties and current metric-based phylogeny.

Given an alphabet $\mathcal{A}$, let $\mathcal{A}^l$ denote the set of all sequences of length $l$ over $\mathcal{A}$. In the case of DNA sequences, $\mathcal{A}=\{A,C,G,T\}$. The \emph{Hamming distance} between two sequences $w_0, w_1\in \mathcal{A}^l$, denoted by $h(w_0,w_1)$, is the number of positions in which the two sequences differ.

It is easy to check that $h$ satisfies the following axioms making it a metric. Namely, for any $w_0,w_1,w_2\in \mathcal{A}^l$
\begin{enumerate}
    \item (identity of indiscernibles) $h(w_0,w_1)=0 \Longleftrightarrow w_0=w_1$.
    \item (symmetry) $h(w_0,w_1)=h(w_1,w_0)$.
    \item (triangle inequality) $h(w_0,w_2)\leq h(w_0,w_1)+h(w_1,w_2)$.
\end{enumerate}

Given a set of sequences in $\mathcal{A}^l$, all pairwise distance information can be encoded in a symmetric matrix, or equivalently, a complete weighted graph, where each node represents a sequence and the weight of the edge represents the Hamming distance between the two corresponding sequences.

\begin{figure}[ht]
\centering
\includegraphics[width=0.6\textwidth]{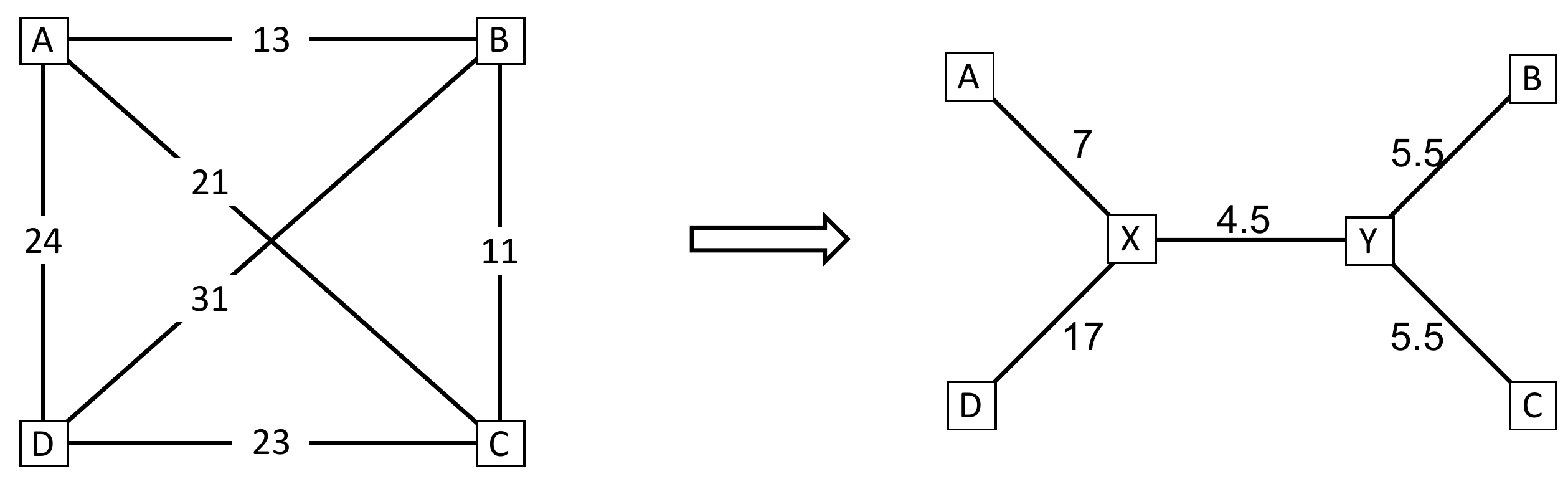}
\caption{LHS: a complete graph that encodes all pairwise dissimilarities among sequences labelled $A$, $B$, $C$ and $D$. The edge label represents the distance between the corresponding sequences. RHS: the corresponding neighbor-joining tree: $X$ and $Y$ were added as interior nodes, and the distance between two original nodes is approximated by the sum of the length of the edges on the unique path the connects these nodes in the tree, e.g. $d'(A,B)=7+4.5+5.5=17$ approximates the original $d(A,B)=13$.}\label{F:tree}
\end{figure}

This metric structure over $\mathcal{A}^l$ allows to infer the phylogenetic relations of a given set of sequences via recursive clustering (i.e. Neighbor-joining, UPGMA, WPGMA etc.~\cite{saitou1987neighbor,sokal1958statistical}) or via optimization (minimum evolution, least squares inference etc.~\cite{fitch1967construction,hendy1982branch}). These relations can then be represented in the phylogenetic tree where the input sequences become the tree's leaves and the internal nodes can then be interpreted as common ancestors of their descendants. The key idea in all such algorithms is that no matter the metric information on the sequences, our target (the phylogenetic tree) is acyclic and spanning. As a result there exists a unique path between any two leaves, i.e.~an unique distance which approximates the original distance, see example~\ref{F:tree}.

In what follows we shall mimic this construction but move beyond pairwise metrics to $k$-wise comparisons ($k$-stances). As a canonical analogue of  the weighted graph, encoding the metric structure, the dissimilarity 
complex will encapsulate the $k$-stances.

\section{k-stances}\label{S:kstances}

As previously noted, one encounters $k$-ary interactions that have no pairwise (Hamming) analogue. In this section we will derive a $k$-spectrum of measurements that capture such higher order dissimilarity relations among multiple sequences. The following Subsections deal with the properties of this measurement and its connection to a particular type of genetic recombination.

We begin by reformulating the Hamming distance between two sequences. Consider the following projections: for any $j\in\{1,\cdots,l\}$, let $f_j:\mathcal{A}^l\xrightarrow{} \mathcal{A}$ where for a sequence $w\in \mathcal{A}^l$, $f_j(w)$ is the letter at position $j$ in $w$. Then for two sequences $w_0,w_1\in \mathcal{A}^l$, each position $j$ contributes one unit to the Hamming distance between the two sequences if and only if $f_j(w_0)\neq f_j(w_1)$.

Accordingly, a position $j$ for the two sequences contributes one unit to the distance, if the number of distinct letters that appear at said position in $w_0$ and $w_1$ respectively, is equal to the number of sequences, i.e. two in this case. Stated this way, the definition of Hamming distance immediately hints at a generalization for any number $k\ge 1$ of sequences as follows

\begin{definition}
Let $k\ge 1$. The \emph{$k$-th order dissimilarity} or \emph{$k$-stance}, of $k$ given sequences each of length $l$, is
given by
$$
d_k:(\mathcal{A}^l)^k \xrightarrow{} \mathbb{N}\quad,\quad
d_k(w_0,\cdots,w_{k-1}):=|\{j\in[[1,l]] : |\bigcup_{i=0}^{k-1}\{f_j(
w_i)\}|=k\}|.
$$
\end{definition}

\begin{figure}[ht]
\centering
\includegraphics[width=0.4\textwidth]{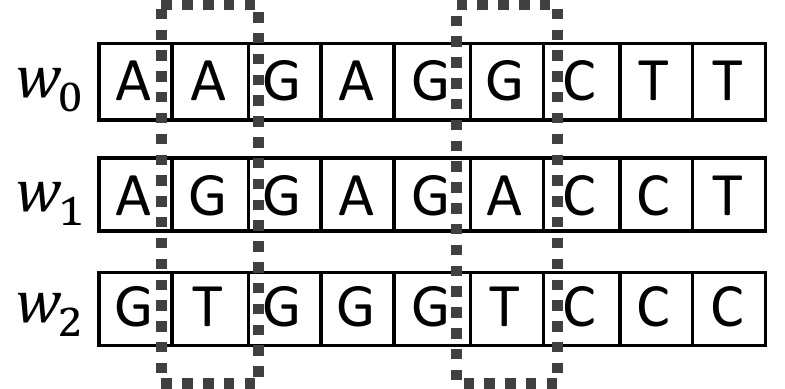}
\caption{$w_0=AAGAGGCTT$, $w_1=AGGAGACCT$ and $w_2=GTGGGTCCC$. Note that at position $2$ and position $6$, all three sequences are mutually distinct. As such $d_3(w_0,w_1,w_2)=|\{j\in[[1,9]] : |\bigcup_{i=0}^{3-1}\{f_j(
w_i)\}|=3\}|=|\{2,6\}|=2$.}\label{F:tristance}
\end{figure}

In other words, the $k$-stance of $k$ given sequences is the number of positions in which the given sequences are all \emph{mutually} distinct, see Figure~\ref{F:tristance}.

\subsection{Basic properties}
\label{S:kstance prop}

By definition, $d_1=l$, i.e. the $1$-stance reproduces the sequence length, while the $2$-stance is exactly the Hamming distance, $d_2=h$. In case of $k>2$, the $k$-stances has the following properties.

\begin{figure}[ht]
\centering
\includegraphics[width=0.6\textwidth]{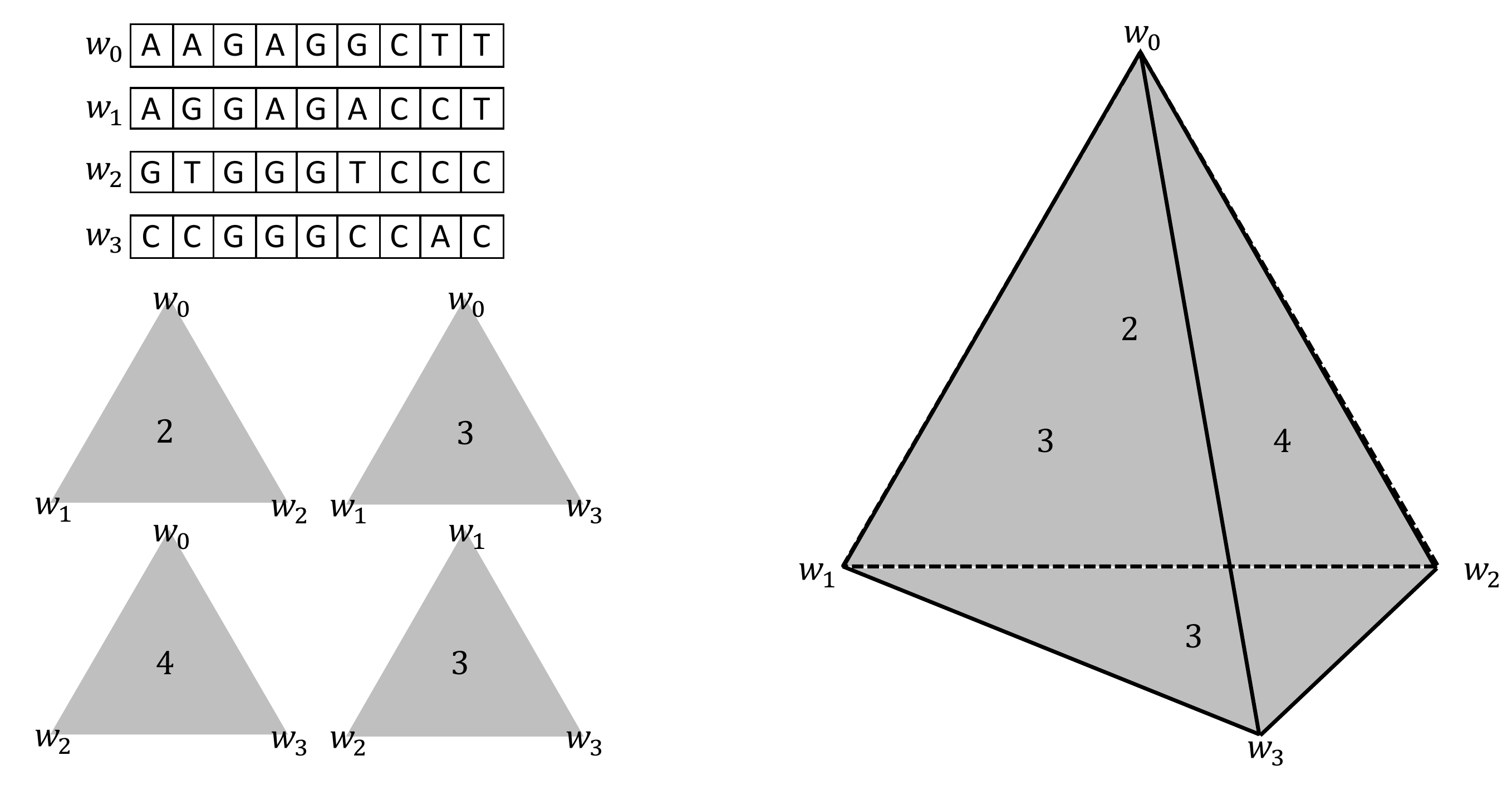}
\caption{An alignment $W=\{w_0=AAGAGGCTT, w_1=AGGAGACCT, w_2=GTGGGTCCC, w_3=CCGGGCCAC\}$, $3$-stances, $d_3(w_0,w_1,w_2)=2, d_3(w_0,w_1,w_3)=3, d_3(w_0,w_2,w_3)=4, d_3(w_1,w_2,w_3)=3$. Note that for instance, $d_3(w_0,w_2,w_3)=4\le 8= 2+3+3=d_3(w_0,w_1,w_2)+d_3(w_0,w_1,w_3)+d_3(w_1,w_2,w_3)$, and this holds for any other permutation as well.}\label{F:inequality}
\end{figure}

\begin{proposition}\label{P:hypermetric}
For any $k>2$ and for any $w_0,\cdots,w_k\in \mathcal{A}^l$, the $k$-stance satisfies
\begin{enumerate}
     \item (implication of indiscernibles) if $w_i=w_{i'}$ for some $0\leq i<i'\leq k-1$ then $d_k(w_0,\cdots,w_{k-1})=0$.
    \item (symmetry) $d_k(w_0,\cdots,w_{k-1})=d_k(w_{\epsilon(0)},\cdots,w_{\epsilon(k-1)})$for any index permutation $\epsilon \in S_k$.
    \item (polyhedron inequality) $d_k(w_0,\cdots,w_{k-1})\leq \sum_{i=0}^{k-1}d_k(w_0,\cdots,\widehat{w_i},\cdots,w_{k})$, where $\widehat{w_i}$ denotes the omission of the $i$-th sequence, See Figure~\ref{F:inequality}.
\end{enumerate}
\end{proposition}

Accordingly, for $k>2$, $d_k$ can be considered as a higher order \emph{pseudometric}~\cite{collatz2014functional}. Note that several generalization of metrics and their properties have been studied in the literature~\cite{klein1998distances, deza1997geometry, sommerville2020introduction}. In particular, in~\cite{klein1998distances} such polyhedron type inequalities appear for certain graph invariants in the context of sums of powers of volumes of $k$-vertices in a graph. This higher dimensional ``volume'' is still derived from pairwise vertex quantities via Cayley-Menger type constructions~\cite{sommerville2020introduction}. In~\cite{deza1997geometry} generalizations of the triangle inequality to hypermetrics are also presented in the context of the cut cone and integer programming, all of which are based on pairwise relations. Dissimilarity, in contrast, entails genuine $k$-interactions.

\begin{proposition}\label{P:dimensionality}
For any $k>2$ and for any $w_0,\cdots,w_{k-1}\in \mathcal{A}^l$, the $k$-stance satisfies
\begin{enumerate}
    \item (dimensional bounding) $d_k(w_0,\cdots,w_{k-1})=0$, for any $k>|\mathcal{A}|$.
    \item (dimensional monotonicity) $d_k(w_0,\cdots,w_{k-1})\le d_{k'}(w_{\iota(0)},\cdots,w_{\iota(k'-1)})$, for any $1\leq k'\leq k$ and any injection $\iota:\{0,\cdots,k'-1\}\xrightarrow[]{}\{0,\cdots,k-1\}$.
\end{enumerate}
\end{proposition}

%We furthermore observe

\subsection{k-stances and genetic recombination}\label{S:kstance recomb}

Genetic recombination can be defined as the exchange of genetic material among multiple sequences and is a key contributor to genetic variation~\cite{rubio2001genetic}. In this section, we focus on a particular class of recombination and discuss its connections to $k$-stances.

\begin{definition}
Let $W=\{w_0,\cdots,w_{k-1}\}\subset\mathcal{A}^l$ be $k$ sequences and fix another sequence $w\in \mathcal{A}^l$. Then $w$ is called a \emph{linear recombinant} of $W$, if for each $j\in\{1,\cdots,l\}$ there exists an $i\in\{0,\cdots,k-1\}$ such that $f_j(w)=f_j(w_i)$.
\end{definition}

Then we obtain

\begin{proposition}\label{P:linrec}
Let $W=\{w_0,\cdots,w_{k-1}\}\subset\mathcal{A}^l$ be $k$ fixed sequences. Then, $d_k(w_0,\cdots,w_{k-1})=0$ if there exists $w\in W$ such that $w$ is a linear recombinant of $W\setminus\{w\}$.
\end{proposition}

\begin{example}
Consider $W=\{w_0=GCTT, w_1=TTCA,  w_2=GCCA \}$. Firstly, $d_1(w_0)=d_1(w_1)=d_1(w_2)=4$, $d_2(w_0,w_1)=4$, $d_2(w_0,w_2)=d_2(w_1,w_2)=2$. Since $w_2=f_1(w_0)f_2(w_0)f_3(w_1)f_4(w_1)$, $w_2$ is a linear recombinant of $W\setminus\{w_2\}$, and as such $d_3(w_0,w_1,w_2)=0$.
\end{example}

Using higher order dissimilarity (as low in dimensionality as $3$-stance) we obtain a more refined description of a sequence set. The following example illustrates this point in providing two sets of sequences, exhibiting the same Hamming distance signature, while differing on the level of $3$-stances.

\begin{example}\label{E:2}
Let $W_0=\{w_0^0=TTCA, w_1^0=CTCG, w_2^0=TTTG\}$ and $W_1=\{w_0^1=CTCG, w_1^1=TTAG, w_2^1=ATTG\}$ be two sets of length four sequences. We have $d_1(w_0^j)=d_1(w_1^j)=d_1(w_2^j)=4$, $d_2(w_0^j,w_1^j)=d_2(w_0^j,w_2^j)=d_2(w_1^j,w_2^j)=2$ for $j=0,1$. However, $d_3(w_0^0,w_1^0,w_2^0)=0\ne 2=d_3(w_0^1,w_1^1,w_2^1)$. Accordingly for $d_1$ and $d_2$, $W_0$ and $W_1$ exhibit identical dissimilarity while their $d_3$-dissimilarities are distinct.
\end{example}

It is worth pointing out that, in the above example, none of the $W_0$-sequences is a linear recombinant of the remaining two, while their $3$-stance is still zero. This indeed shows that the implication of indiscernibles for $k$-stances with respect to linear recombinants is a sufficient but not a necessary condition.

\section{The dissimilarity complex}\label{S:dissimilarity complex}

To understand the extra arithmetic information encapsulated within the dissimilarity complex, we recall the notion of simplicial complexes. We adopt here a data-centric point of view and eschew abstract topological and category theoretical considerations. 

\begin{figure}[ht]
\centering
\includegraphics[width=0.6\textwidth]{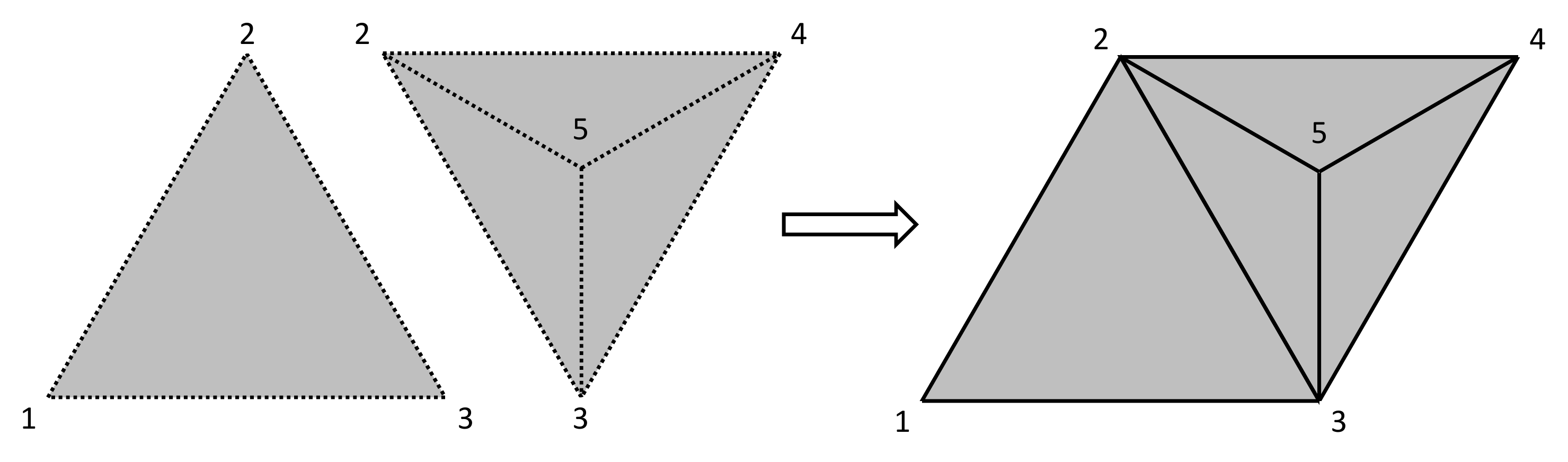}
\caption{A simplicial complex $(W,\wedge)$, where $W=\{1,2,3,4,5\}$ with  $\wedge\{1,2,3\}$ and $\wedge\{2,3,4,5\}$. These two polytopes share the one dimensional line segment sub-polytope $\{2,3\}$. Gluing produces the geometric realization of $(W,\wedge)$.}\label{F:complex}
\end{figure}

Suppose we are given a discrete, finite set of data points $W$ as measurements of a system, and suppose that among the points in this data set there exists a relation denoted by $\wedge$. This relation might model an intrinsic dependency in the system that manifests in the measurements in $W$. Suppose that the relation $\wedge$ had the property that for any subsets $W''\subseteq W'\subseteq W$ we have that $\wedge W'\implies \wedge W''$. Namely, if a subset of elements in $W$ are in the $\wedge$ relation, then any subset of those elements are in the $\wedge$ relation as well. We shall also assume  that for each individual measurement $w\in W$ we have $\wedge\{w\}$. Then, the combinatorial structure consisting of the totality of such $\wedge$-satisfying subsets is called the simplicial complex of the data set $W$ under the relation $\wedge$ and is denoted by $(W,\wedge)$. A single such element $W'\in (W,\wedge)$ is called a $|W'|-1$-dimensional simplex. This complex can be organised into a topological object by embedding each simplex $W'\in (W,\wedge)$ into a $(|W'|-1)$-dimensional $\mathbb{R}$-linear (Euclidean) polytope, and gluing these polytopes along their common faces, see Figure~\ref{F:complex}.

We shall use this perspective to build the dissimilarity complex by letting $W$, the data set, be the sequences in a (genetic) alignment, with the relation $\wedge:=$\emph{``mutual dissimilarity at at least one position''}. Furthermore, we keep track of the degree of dissimilarity for each such $k-1$-simplex via the $k$-stance among its constituent sequences.

\subsection{Construction of the dissimilarity complex}\label{S:construction of dissimilarity complex}

We are now in position to formally introduce the natural mathematical structure encapsulating higher order dissimilarity information among a collection of sequences.

\begin{definition}
An \emph{alignment} $W=[w_1,\cdots,w_n]$, $w_i\in\mathcal{A}^l$, is a finite ordered tuple of sequences of equal length $l$ over an alphabet $\mathcal{A}$. We can view $W$ as a matrix whose entries are letters in $\mathcal{A}$ with $w_i$ being the sequence (ordered tuple of letters) in the $i$-th row of $W$. Furthermore, the $j$th column of $W$ is the ordered tuple $[f_j(w_1),\cdots,f_j(w_n)]$. The integer $l$ is called the length of the alignment (the number of columns) while the integer 
$n$ is called the size of the alignment (the number of sequences).
\end{definition}

Given an alignment, its dissimilarity signature is expressed via a weighted simplicial complex, defined as follows.

\begin{definition}
Let $W=[w_1,\cdots,w_n]$, $w_i\in\mathcal{A}^l$, be a given alignment and let $k\ge 0$ be fixed. A \emph{simplex} of dimension $k$ over $W$ is a $k+1$-subset of $W$, $\sigma=\{w_0,\cdots,w_k\}$ of sequences from $W$, such that $d_{k+1}(w_0,\cdots,w_k)>0$. We denote by $K_k(W)$ the set of all possible $k$-simplices over $W$, and set $X(W)=\cup_{k\ge 0}K_k(W)$. Let $R\supset \mathbb{Z}$ be a discrete valuation ring with uniformizer $\pi$. Let
$$v_W: X(W) \xrightarrow{} R\quad,\quad v_W(\sigma=[w_0,\cdots,w_k])=\pi^{d_{k+1}(w_0,\cdots,w_k)}.$$
Then, $v_W$ is called the \emph{weight function} associated to $X(W)$ and we call the pair $(X(W),v_W)$ the \emph{dissimilarity complex} of $W$.
\end{definition}

We can think of the discrete valuation ring as being a polynomial ring in the transcendent variable $\pi$ with rational coefficients, where we use the powers of $\pi$ to express the weights of simplices.

Given a dissimilarity complex, we can construct its \emph{geometrical realization} by constructing the geometrical realizations at each $k$ dimension and then integrating them via gluing along common faces, see Example~\ref{E:geometrical realization}.

\begin{figure}[ht]%
\centering
\includegraphics[width=0.9\textwidth]{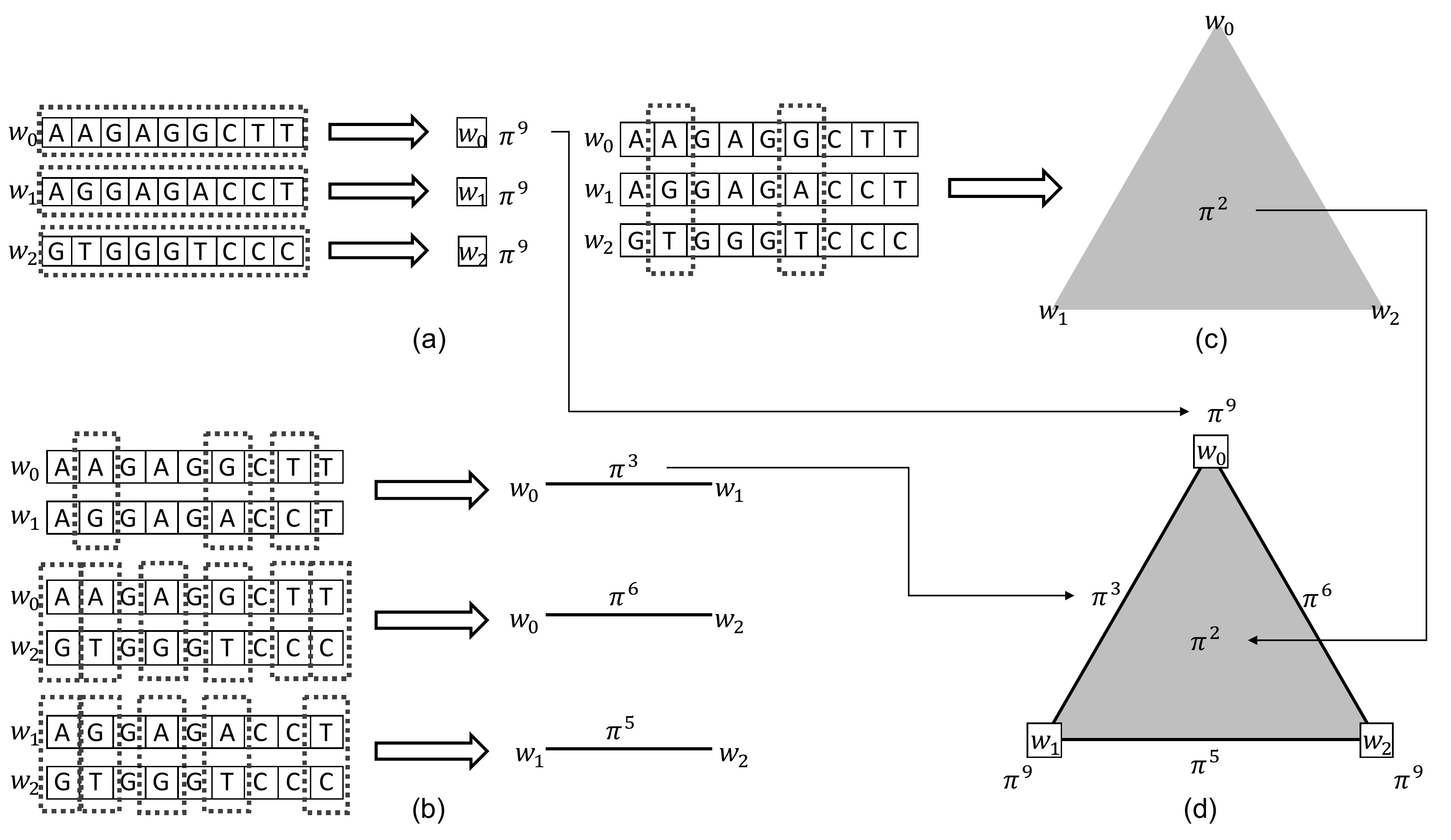}
\caption{The geometric realizations at different dimensions and their integration of $X([w_0=AAGAGGCTT, w_1=AGGAGACCT, w_2=GTGGGTCCC])$ with $K_0$ in (a), $K_1$ in (b), $K_2$ in (c) and the integration into $(X(W),v_W)$ in (d).}\label{F:geometric}
\end{figure}

\begin{example}\label{E:geometrical realization}
Let $W=[w_0=AAGAGGCTT, w_1=AGGAGACCT, w_2=GTGGGTCCC]$. We can construct the geometrical realization of $(X(W),v_W)$ by first constructing its $0$-simplices (see Figure~\ref{F:geometric}(a)), $1$-simplices (see Figure~\ref{F:geometric}(b)) and $2$-simplices (see Figure~\ref{F:geometric}(c)). The geometrical realization of $(X(W),v_W)$ is then obtained by integrating all $k$-simplices via gluing (see Figure~\ref{F:geometric}(d)).
\end{example}

An immediate motivation for this construction is that more information can be encoded when we lift from weighted graphs to the weighted complex structure, see Example~\ref{E:more}. We are now in position to identify information not present in the graph-- which is only the so called $1$-skeleton of the complex.

\begin{figure}[ht]%
\centering
\includegraphics[width=0.6\textwidth]{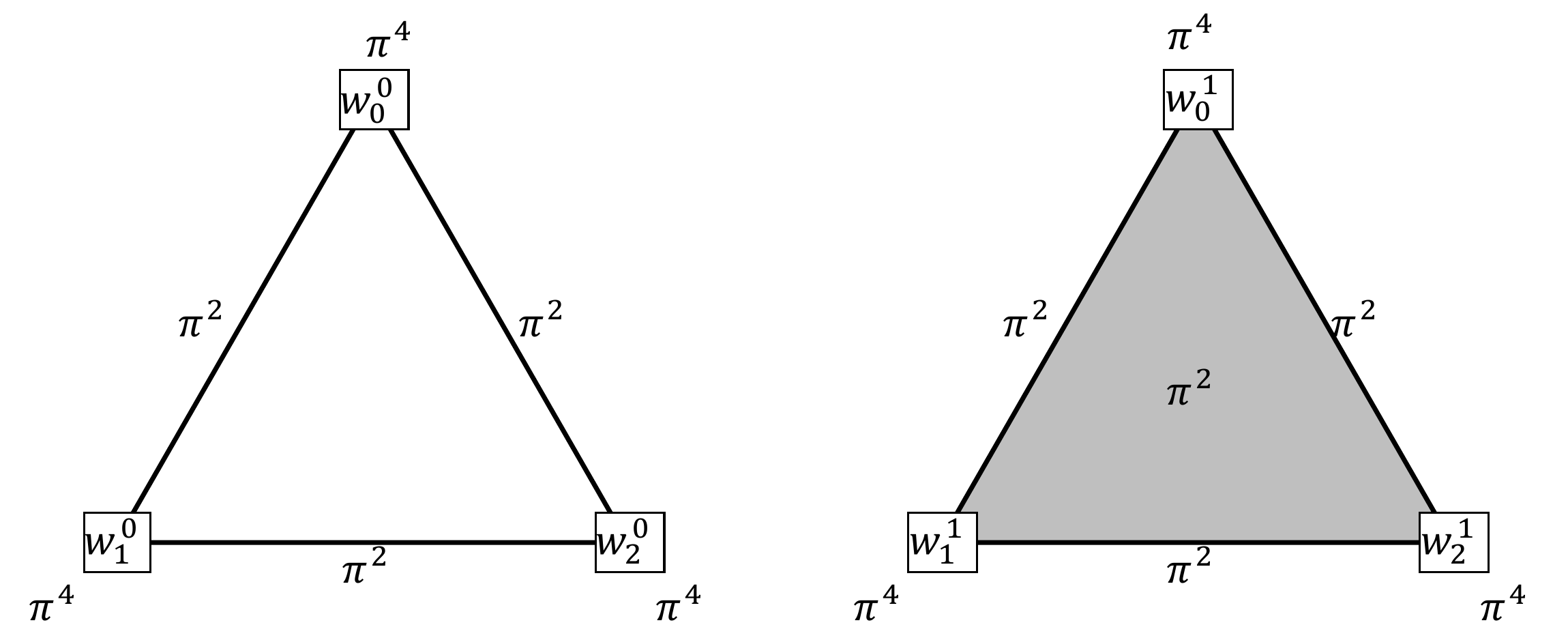}
\caption{Dissimilarity complexes corresponding to $W_0=[w_0^0=TTCA, w_1^0=CTCG, w_2^0=TTTG]$ and $W_1=[w_0^1=CTCG, w_1^1=TTAG, w_2^1=ATTG]$, with their respective weights. Note that $X(W_0)$ is an ``empty'' triangle while $X(W_1)$ is ``filled''.}\label{F:triangle}
\end{figure}

\begin{example}\label{E:more}
Consider the dissimilarity complexes associated to the alignments in Example~\ref{E:2}. For $W_0=[w_0^0=TTCA, w_1^0=CTCG, w_2^0=TTTG]$, We have $d_1(w_0^0)=d_1(w_1^0)=d_1(w_2^0)=4\neq 0$, hence $K_0(W_0)=\{[w_0^0], [w_1^0], [w_2^0]\}$ and since $d_2(w_0^0,w_1^0)=d_2(w_0^0,w_2^0)=d_2(w_1^0,w_2^0)=2\neq 0$, we also have $K_1(W_0)=\{[w_0^0,w_1^0],[w_0^0,w_2^0],[w_1^0,w_2^0]\}$. Finally  $d_3(w_0^0,w_1^0,w_2^0)=0$, thus $K_2(W_0)=\varnothing$, see Figure~\ref{F:triangle} (LHS). For $W_1=[w_0^1=CTCG, w_1^1=TTAG, w_2^1=ATTG]$ we have $K_0(W_1)\cong K_0(W_0)$ and $K_1(W_1)\cong K_1(W_0)$ however $d_3(w_0^1,w_1^1,w_2^1)=2\neq 0$ yields $[w_0^1,w_1^1,w_2^1]\in K_2(W_1)$, see Figure~\ref{F:triangle} (RHS).
\end{example}

$(X(W),v_W)$ represents an augmentation of classical simplical complexes and in particular a generalization of weighted graphs. In addition, $(X(W),v_W)$ has some nice combinatorial properties that facilitate the study of a weighted version of homology as detailed in Section~\ref{S:the weighted homology of the dissimilarity complex}.

\begin{proposition}\label{P:wdef}
$X(W)$ is a simplicial complex that is bounded in dimension, namely, for any $\sigma\in X(W)$, $\dim(\sigma)\le|A|-1$.
\end{proposition}

\begin{proposition}\label{P:wsc}
Let $\sigma\in X(W)$ be a $k$-simplex and let $\tau\subseteq \sigma\in X(W)$ be a $k'$ face of $\sigma$. Then we have $v_W(\sigma)|v_W(\tau)$.
\end{proposition}

\begin{definition}\label{shuffles}
Let now $\epsilon\in S_n$ be a permutation on $\{1,\cdots,n\}$. The ordered tuple $$W_\epsilon:=[w_{\epsilon(1)},\cdots,w_{\epsilon(n)}],$$ is called a \emph{row shuffle} of $W$. 

Let $\omega\in S_l$ be a permutation on $\{1,\cdots,l\}$. The ordered tuple 
$$W^\omega:=[w_1^\omega,\cdots,w_l^\omega],$$ with $$w_i^\omega=[f_{\omega^{-1}(1)}(w_i),\cdots, f_{\omega^{-1}(l)}(w_i)],$$ for all $1\le i\le l$, is called a \emph{column shuffle} of $W$.
\end{definition}

\begin{proposition}\label{P:shuffle}
For any alignment $W$ and any pair of row and column shuffles $(\epsilon,\omega)\in S_n\times S_l$, we have $(W_\epsilon)^\omega=(W^\omega)_\epsilon$.
Furthermore denoting $W_\epsilon^\omega:=(W_\epsilon)^\omega=(W^\omega)_\epsilon$ we have that, for any pair of row and column shuffles $(\epsilon,\omega)\in S_n\times S_l$, $$(X(W),v_W)\cong (X(W_\epsilon^\omega),v_{W_\epsilon^\omega}).$$
\end{proposition}

Proposition~\ref{P:wsc} shows that $(X(W),v_W)$ is amenable to construct weighted homology by means of a novel 
boundary operator, compatible with the weight function, see Section~\ref{S:the weighted homology of the dissimilarity complex}. Proposition~\ref{P:shuffle} shows that the dissimilarity complex has nice symmetry properties.

\subsection{Weighted homology}\label{S:the weighted homology of the dissimilarity complex}

Passing from graphs to simplicial complexes not only provides us with the degree of freedom to encode additional information, but it also enables us to study a multiple sequence alignment from a novel mathematical perspective.
Any simplicial complex gives rise to a topological space. Studying topological properties of the dissimilarity complex 
enhances our conceptual understanding of the multiple sequence alignment itself.

In algebraic topology, simplicial homology is a useful tool for the study of features of a simplicial complex. It comes 
about as a sequence of abelian groups $H_1,\cdots H_n,\cdots$, one for each dimension, whose structures yield surprising information (invariants with respect to continuous deformations) about the space in question, such as the structure of its $k$-dimensional holes and its orientability (geometric torsion). This information is of key relevance and to date, dynamically tracking the birth and death of generators of these homology groups is an integral part of topological data analysis in the guise of Persistent Homology~\cite{zomorodian2005computing}.

The dissimilarity complex constitutes a simplicial complex with additional weight information. By coherently integrating this weight information into a new homology theory that mimics the classical case, we can study the augmented arithmetic version of its topology. In this case, its torsion encodes $k$-stance level information among the sequences, and thus we can gain more insight about the structure of the alignment the dissimilarity complex is modeling.

Given an alignment $W$, let $(X,v)=(X(W),v_W)$ be its corresponding 
weighted dissimilarity complex. Let $C_{n,R}(X)$ denote the free $R$-module generated by all $n$-simplices in $X$, with $R$ being the co-domain ring of the simplex weight function. Setting a \emph{simplicial ordering}~\cite{hatcher2005algebraic}, namely a linear order on the $0$-simplices, we can now consider a simplex $\sigma$ as an ordered tuple of sequences instead of a set. This allows one to define
$$
\partial^v_n:C_{n,R}(X)\to C_{n-1,R}(X)\quad,\quad\partial^v_n(\sigma)=\sum_{i=0}^n  \frac{v(\hat{\sigma}_i)}{v(\sigma)}\cdot (-1)^i\hat{\sigma}_i,
$$
where the face $\hat{\sigma}_i\subset\sigma$ is obtained by dropping the $i$th position in $\sigma$. We have $v(\sigma)$ divides $v(\hat{\sigma}_i)$ and as a result, $\partial^v_n$ is a well defined $R$-module homomorphism.
Note that
$\frac{v(\hat{\sigma}_{i,j})}{v(\hat{\sigma}_i)}\cdot \frac{v(\hat{\sigma}_i)}{v(\sigma)}=
\frac{v(\hat{\sigma}_{j,i})}{v(\hat{\sigma}_j)}\cdot \frac{v(\hat{\sigma}_j)}{v(\sigma)}
$, hence we obtain $\partial^v_{n-1}(\partial^v_n(\sigma))=0$. In view of this, $\partial^v_n$ is a boundary map and we can define a homology theory accordingly~\cite{bura2021weighted,dawson1990homology,ren2018weighted}, namely $H^v_{n}(X)=\text{\rm Ker}(\partial^v_n)/\text{\rm Im}(\partial^v_{n+1})$ denote the weighted homology modules of $(X,v)$. Furthermore, it is a well known result that the weighted homology modules are in fact independent of our initial choice of simplicial order~\cite{hatcher2005algebraic}.

\begin{proposition}\label{P:dimensionality 2}
Let $W$ be an alignment over an alphabet $\mathcal{A}$ and let $(X(W),v_W)$ be its corresponding weighted dissimilarity complex. Then $H^v_{k}(X(W))=0$, for all $k\geq |\mathcal{A}|$.
\end{proposition}

\begin{proposition}\label{P:shuffle 2}
Let $W$ be an alignment consists of $n$ sequences of length $l$. For any pair $(\epsilon,\omega)\in S_n\times S_l$, and any $k\in \mathbb{N}$, we have $H^{v_W}_{k}(X(W))\cong H^{v_{W_{\epsilon}^{\omega}}}_{k}(X(W_{\epsilon}^{\omega}))$.   
\end{proposition}

\section{The single column case}\label{S:mathematics of single column}
In this section we present relation between the $k$-stances for different $k$ for alignments of length one. 
Furthermore we compute their weighted homology. To this end, let $W=[w_0,\cdots,w_n]$ be an alignment of length one and size $n$.

\begin{figure}[ht]%
\centering
\includegraphics[width=0.6\textwidth]{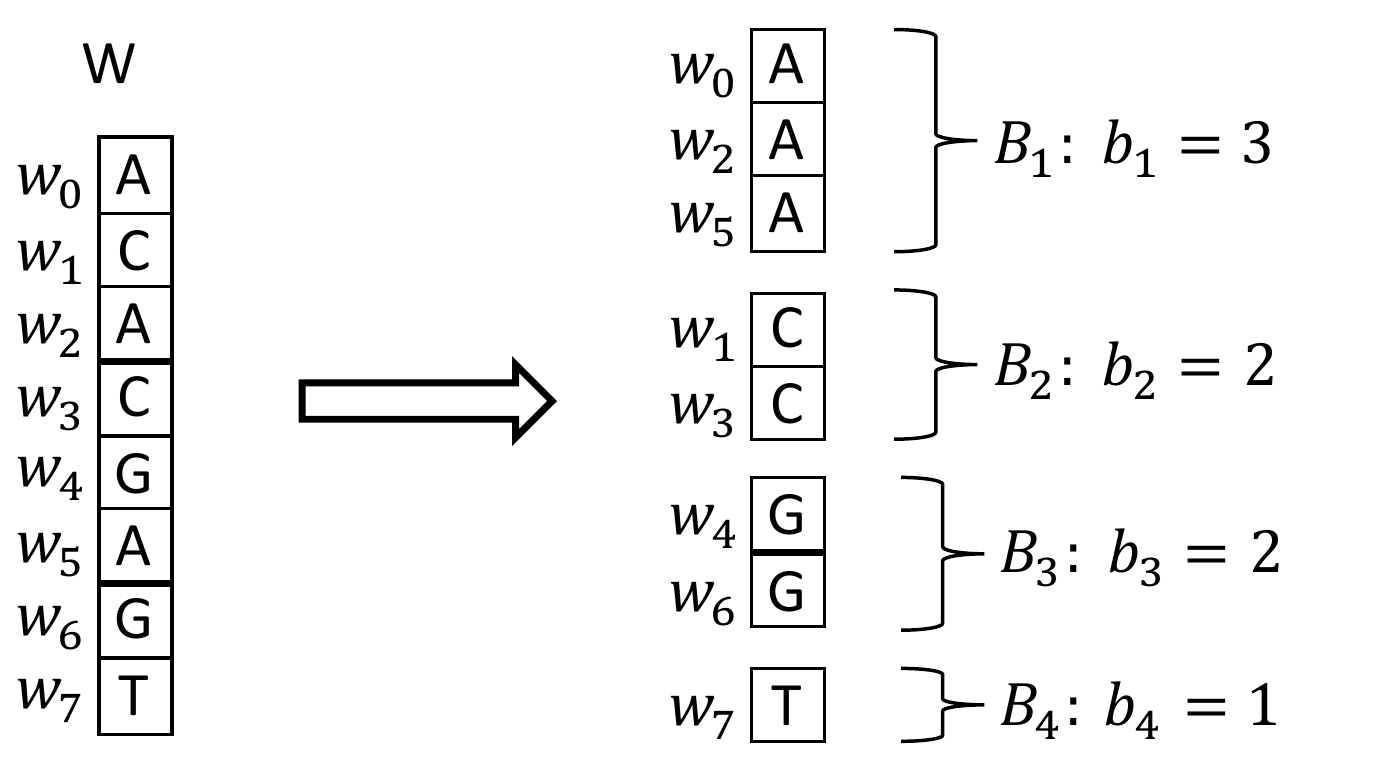}
\caption{The partition of a single column alignment $W$ into bins.}\label{F:bin}
\end{figure}

We can organize $W$ via bins $W=\dot\cup_{i=1}^s B_i=\dot\cup_{i=1}^s\{w \in W|f_1(w)=a_i\}\ne\varnothing$ where 
$s$ is the number of distinct letters that appear in $W$'s column and we let $b_i=|B_i|$ for all $1\leq i\leq s$ be the size 
of $B_i$, i.e. the multiplicity in $W$'s column of the letter $a_i$, see Figure~\ref{F:bin}.

\begin{figure}[ht]%
\centering
\includegraphics[width=0.6\textwidth]{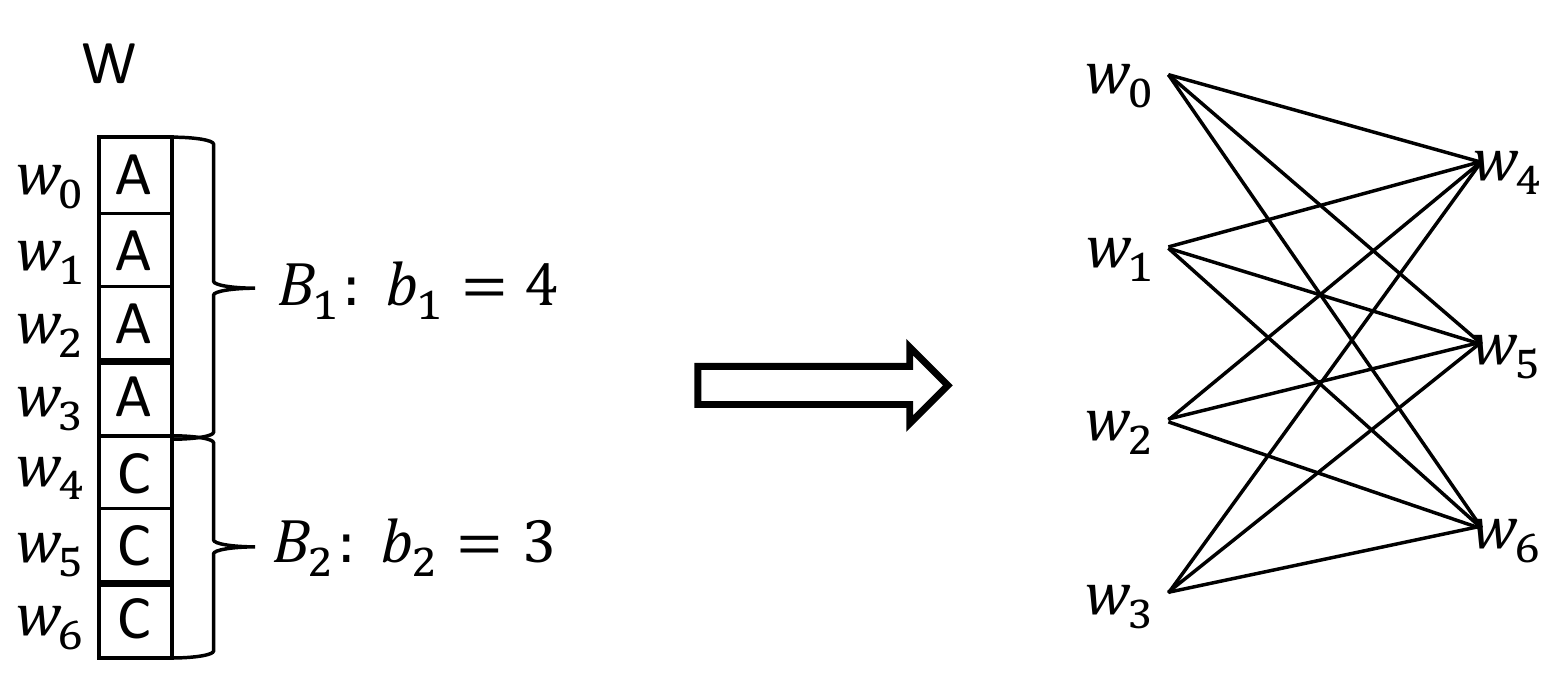}
\caption{A single column alignment with only two bins and its dissimilarity complex, the complete bipartite graph $K_{4,3}$.}\label{F:k-partite}
\end{figure}

Given the bin partitioning of $W$, it is easy to see that any simplex $\sigma\in X(W)$ has weight $v(\sigma)=\pi^1=\pi$. Furthermore $X(W)$ is a \emph{pure} simplicial complex as all of its maximal simplices are of dimension $s-1$. By construction, each $s-1$-simplex is obtained by picking one sequence ($0$-simplex) from each of the $s$ bins. Therefore, $X(W)$ is a \emph{complete $k$-partite simplicial complex}, which is a natural generalization of the complete bipartite graph. Note that in the case of $s=2$, $X(W)$ is precisely the classical complete bipartite graph $K_{b_1,b_2}$, see Figure~\ref{F:k-partite}.

\begin{definition}
The total $k$-stance contribution in $W$ is defined to be

$$
c_k=\sum_{Y\subseteq W, |Y|=k} d_k(y_0,\cdots,y_{k-1}),
$$

where the sum is taken over all size $k$ subsets $Y=\{y_0,\cdots,y_{k-1}\}\subset W$. We integrate this information over all $k$ into a polynomial in indeterminate $x$ called the dissimilarity polynomial of $W$

$$D_W(x)=x^s+\sum_{k=1}^s (-1)^kc_kx^{s-k}.$$

\end{definition}

\begin{theorem}\label{T:Vieta}
Let $W$ be a single column alignment. The size of each bin of $W$ is a root of $W$'s dissimilarity polynomial, and this polynomial has no other roots.
\end{theorem}

\begin{example}
Let $W$ be the single column alignment shown in Figure~\ref{F:bin}. We have $b_1=3$, $b_2=b_3=2$ and $b_4=1$. Furthermore, we have $c_1=8$, $c_2=23$, $c_3=28$ and $c_4=12$. Then $D_W(x)=x^4-8x^3+23x^2-28x+12=(x-3)(x-2)^2(x-1)$.
\end{example}

\begin{theorem}\label{T:bins}
Let $W$ be a single column alignment, let $(X,v):=(X(W),v_W)$ be its corresponding weighted simplicial complex and denote by $b=\prod_{i=1}^{s}|b_i-1|$. Then, all homology modules $H^v_{k}(X)$ are free and furthermore
\begin{itemize}
    \item $a)$ $H^v_{0}(X)=R$,
    \item $b)$ $H^v_{s-1}(X)=R^b$
    \item $c)$ $H^v_{k}(X)=0$ for any $k>0,k\ne s-1$.
\end{itemize}
\end{theorem}

\begin{example}
Let $W$ be the single column alignment shown in Figure~\ref{F:k-partite}. We have $b_1=4$ and $b_2=3$. Then $H_0^v(X)=R$, $H_1^v(X)=R^{(4-1)(3-1)}=R^6$ and $H_{n\ge 2}^v(X)=0$.
\end{example}

\section{Dissimilarity and k-stances of multi-column sequence alignments}\label{S:multi column case}

For general alignments we have at present no analytical (closed form) expression connecting its $k$-stances and 
weighted homology modules in terms of the bins of its various columns. A way of piecing together column information 
inductively is currently under investigation and the idea here would be to employ some version of Mayer-Vietoris sequences for weighted complexes. 
However, $k$-stance statistics as well as the modules of weighted homology can be computed, effectively. We have developed a framework for computing weighted homology and can provide a link to a free underlying software module created for this purpose (\href{https://biocomplexity.virginia.edu/institute/divisions/mathematical-biocomplexity}{software module}). In the following, we shall illustrate that both $k$-stances and weighted homology provide new insights into aligned genetic data and reveal biologically relevant features of said alignments.

In Subsection~\ref{S:k-stance statistics multi column} we present case studies for SARS-CoV-2 and for H1N1, 
respectively, where $k$-stance signatures are seen to reflect distinct phases in the evolution of these pathogens in 
the human population. In Subsection~\ref{S:homology multi column} we illustrate connections between the structure of the weighted homology modules and the $k$-stances present in the alignments.

\subsection{k-stance statistics}
\label{S:k-stance statistics multi column}
In this Subsection we present two case studies that illuminate the usage of higher order $k$-stance statistics to infer biologically relevant information on viral population dynamics.

Case study $1$: SARS-CoV-2

\begin{figure}[ht]%
\centering
\includegraphics[width=0.9\textwidth]{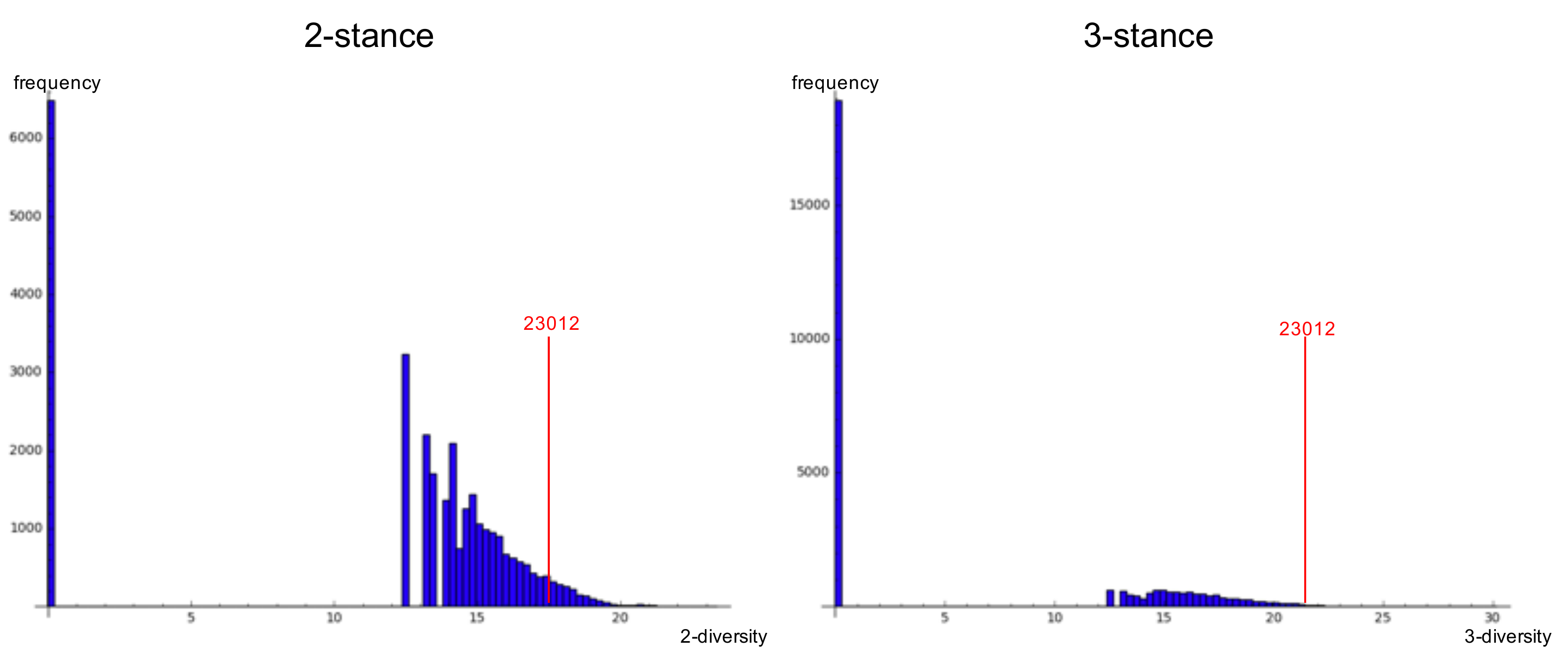}
\caption{Histograms of site $k$-stance distribution ($k=2,3$). $x$-axis: $Log$($k$-stance+1), $y$-axis: frequency in each bin. The red line marks the bin containing site 23012 corresponding to the well studied mutations E484Q and E484K.}\label{F:D614G}
\end{figure}

The multiple sequence alignment considered here is comprised of all SARS-CoV-2 genomes submitted to GISAID \cite{shu2017gisaid} prior to 2021-01-11. This amounts to $254148$ sequences, each exhibiting $29903$ aligned sites.
For each site, we computed its total $2$-stance and $3$-stance contribution respectively (i.e. the total number of pairs and the total number of triplets that are mutually distinct), where the gap symbol was not accounted as a distinguished symbol to any of the contributions computed. We now partition the set of logarithms of these numbers (shifted by 1 for technical reasons) into 100 bins of the same width and plot their corresponding histograms (bin vs frequency), see Figure~\ref{F:D614G}.

The $2$-stance and the $3$-stance distributions contain approx.~$6000$ and $20000$ in the zero-th bin, respectively. In any other bin the two distributions also differ significantly, the $2$-stance exhibits a sharp decay in frequency while the $3$-stance remains relatively flat with only a slight decay. Having a closer look at the polymorphic 
site $23012$ mentioned in the Introduction, corresponding to the E484Q and E484K mutations, we find rank $327$ for $2$-stance and rank $19$ for $3$-stance. This suggests that  the $3$-stance measurement provides a higher signal to noise ratio. Note that the non wild type fraction of sequences in the aggregated population is less than $0.12\%$. In other words, the $3$-stance is highly sensitive and can facilitate early VoI/VoC detection.
    
Case study $2$: H1N1

\begin{figure}[ht]%
\centering
\includegraphics[width=0.6\textwidth]{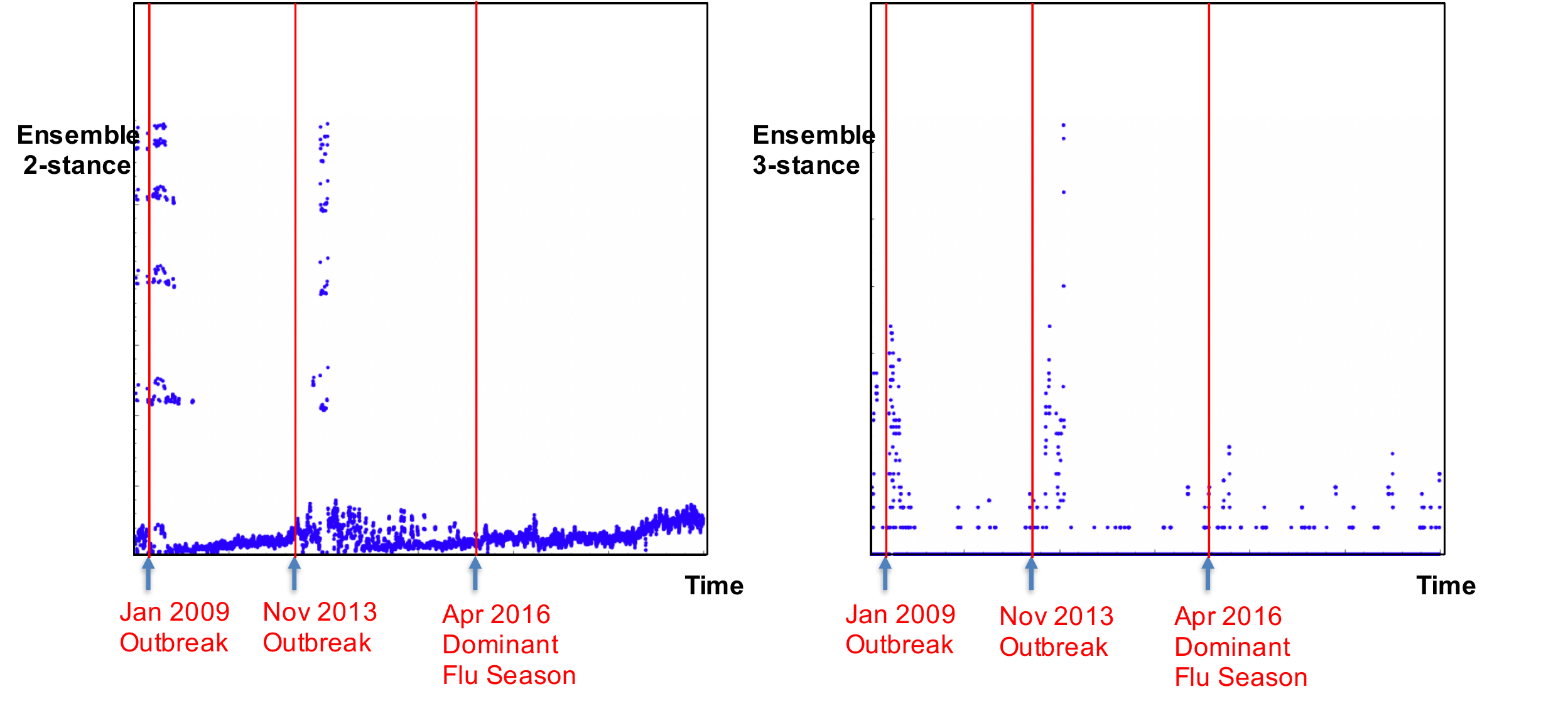}
\caption{Time evolution of the ensembles of $2$-stance and $3$-stance in a sliding window of 100 sequences across a temporally ordered alignment of  H1N1 flu data from 2009 to 2018.}\label{F:flu}
\end{figure}

We study $2$-stances and $3$-stances within a sliding window of 100 sequences across a temporally ordered alignment of GISAID H1N1 flu data from 2009 to 2018. The $y$-axis represents the sum over all possible $k$-stances ($k=2,3$) which we refer to as the ensemble of $2$-stance and $3$-stance, respectively, for each window as time evolves
in the $x$-axis, see Figure~\ref{F:flu}.

$2$- and $3$-stances capture the two outbreaks (Jan 2009 and Nov 2013) of the virus and we speculate that the peaks in this dissimilarity signal appear due to the virus' genetic variation being elevated as it explores its fitness landscape. However, note that the Apr 2016 flu season that exhibited a change of the dominant strain is not captured by $2$- but $3$-stances. We stipulate that this is the case because $3$-stances exhibit higher signal to noise ratio than the $2$-stance.

\subsection{Multi column weighted homology}\label{S:homology multi column}

~\cite{li2022homology} provides structure theorems for the weighted homologies of arbitrary weighted complexes (not necessarily arising from dissimilarity) relating simplicial homology with coefficients in certain valuation rings to weighted simplicial homology. The idea being here is to create a homomorphic image of the ``known'' homology into the ``unknown'' homology and then to study the quotient via homological algebra. The concepts developed in the process suggest employing a version of Nakayama's Lemma~\cite{nakayama1951remark} to reduce the coefficients controlling this quotient down to rational numbers. This enables very fast computation of all weighted homology modules (\href{https://biocomplexity.virginia.edu/institute/divisions/mathematical-biocomplexity}{software module}).

The weighted homology modules of the dissimilarity complex exhibit non-trivial torsion, which genuinely stems from $k$-stances and reflects interesting features about the structure of the alignment itself. 
We present two pertinent examples that allude to this fact:

\begin{example}
Consider the alignment $W=[w_0=AGCTTT, w_1=ATTCAA,  w_2=AGCCAA]$. Firstly, we have $d_1(w_0)=d_1(w_1)=d_1(w_2)=6$. Then we have $d_2(w_0,w_1)=5$, $d_2(w_0,w_2)=3$ and $d_2(w_1,w_2)=2$. Finally, we have $d_3(w_0,w_1,w_2)=0$. Since the maximum dimension of $X(W)$ is one, we have two nontrivial weighted homology modules, namely, $H_1^v(X(W))=R$ and $H_0^v(X(W))=R\oplus R/\pi \oplus R/\pi^3$. Note that $H_1^v(X(W))$ has free rank one, and this is due to the fact that $w_2$ is a linear recombinant of $w_0$ and $w_1$,  namely $w_2=f_1(w_0)f_2(w_0)f_3(w_0)f_4(w_1)f_5(w_1)f_6(w_1)$. Furthermore, $H_0^v(X(W))$ has free rank one and two torsion components. The first torsion component $R/\pi=R/\pi^{(6-5)}$ corresponds to the largest $2$-stance among the three sequences and the second torsion component $R/\pi=R/\pi^{(6-3)}$ corresponds to the second largest $2$-stance among the three sequences.
\end{example}

\begin{example}
Let $W=[w_0=AAGAGGCTT, w_1=AGGAGACCT, w_2=GTGGGTCCC]$. Firstly, we have $d_1(w_0)=d_1(w_1)=d_1(w_2)=9$. Then we have $d_2(w_0,w_1)=3$, $d_2(w_0,w_2)=6$ and $d_2(w_1,w_2)=5$. Finally, we have $d_3(w_0,w_1,w_2)=2$. Since the maximum dimension of $X(W)$ is two, we have $H_k^v(X(W))=0$, for all $k\geq 3$. In fact, we have $H_2^v(X(W))=0$, $H_1^v(X(W))=R/\pi$ and $H_0^v(X(W))=R\oplus R/\pi^3 \oplus R/\pi^4$. Since $H_1^v(X(W))=R/\pi$ are full torsion, none of $w_0$, $w_1$ or $w_2$ is a linear recombinant of the remaining two. Furthermore, the torsion $R/\pi=R/\pi^{(3-2)}$ corresponds to the difference between the $3$-stance and the minimum pairwise $2$-stance among the three sequences.     
\end{example}

\section{Discussion}\label{S:discussion}

In this paper we introduce the notion of higher order dissimilarities, naturally generalizing the concept of Hamming distance. We have shown that such dissimilarities emerge within alignments of viral sequences and that these are not independent of each other. In fact we give explicit formulae for these dependencies in specific instances. We can thus conclude that, in 
case of genetic sequences and the underlying four letter alphabets, there is more information than is reflected by Hamming distance alone by considering $3$- and $4$-stances. It is therefore noteworthy that all the information we currently obtain is based on or derived from Hamming distance.

We then provide a mathematical context for these hyper-distances by means of the dissimilarity complex. Here $k$-stances manifest as weights of certain simplices. To be concrete, simplices are comprised of $k$-sequences that exhibit  in at least one site a $k$th order polymorphism and the weight of this simplex is the actual number of the sites exhibiting such $k$th order polymorphisms. The weighted complex homology can be readily computed via weighted homology~\cite{li2022homology} and in case of sequence alignments, an inductive computation by means of patching the complex column by column that is  based on the single column case--which we compute here--is currently under investigation. 

As for future work, along the lines of constructing the phylogenetic tree within a complete weighted graph of an alignment, we work on constructing a ``tree-analogue'' within the Dissimilarity Complex of the given alignment.
This ``phylogenetic complex'' generalizes the well known phylogenetic tree. It is natural then to ask what sort of properties such a derived object should possess:

Clearly a tree is an acyclic graph. In homological terms, when regarded as a one dimensional complex, it exhibits trivial
homology in dimension one. It is therefore natural to assume that the phylogenetic complex should be homologically trivial across all higher dimensions.

Tree edges form a maximally independent set, i.e.~including any so called ``closing'' edges we obtain cycles and
distances of these closing edges are approximated by the involved tree edges. It is natural then to require that the phylogenetic complex obey similar properties, again across all dimensions. Namely, it should be able to approximate the weight of any added simplex across all dimensions.

The phylogenetic complex will inevitably include higher order, pseudometric information and arises as a result of an optimization process that is fundamentally different from clustering. This is clear since the very notion of clustering is based on pairwise relations.

\section{Proofs}\label{S:proofs}

Proposition~\ref{P:hypermetric}.
\begin{proof}
Implication of indiscernibles: if $w_i=w_{i'}$ for some $0\le i<i'\le k-1$, then $f_j(w_i)=f_j(w_{i'})$ for any $j\in\{1,\cdots,l\}$ and the claim follows by definition of $d_k$.

Symmetry: since $\bigcup_{i=0}^{k-1}\{f_j(w_i)\}=\bigcup_{i=0}^{k-1}\{f_j(w_{\epsilon(i)})\}$ for any $\epsilon\in S_k$, the claim follows by definition of $d_k$.

Polyhedron inequality: Let $I:\mathcal{A}^k\xrightarrow{} \{0,1\}$ be an indicator function for which $I(f_j(w_0),\cdots,f_j(w_{k-1}))=1$ if $|\bigcup_{i=0}^{k-1}\{f_j(w_i)\}|=k$ and $I(f_j(w_0),\cdots,f_j(w_{k-1}))=0$ otherwise. We note then that
$$d_k(w_0,\cdots,w_{k-1})=\sum_{j=1}^l I(f_j(w_0),\cdots,f_j(w_{k-1})).$$
It suffices then to show that $I$ satisfies the polyhedron inequality. Furthermore, since $I$ is always non-negative, it suffices to only consider the case $I(f_j(w_0),\cdots,f_j(w_{k-1}))=1$. If $f_j(w_k)\ne f_j(w_i)$ for all $0\leq i\leq k-1$, then $I(f_j(w_0),\cdots,\widehat{f_j(w_i)},\cdots,f_j(w_k))=1$ for any $0\leq i\leq k-1$. In this case the polyhedron inequality holds for $I$. The other possibility is that $f_j(w_k)=f_j(w_{i^*})$ for some distinguished $0\leq i^*\leq k-1$. But then, $I(f_j(w_0),\cdots,\widehat{f_j(w_{i^*})},\cdots,f_j(w_k))=1$ which still implies the claim for $I$, completing the proof.
\end{proof}

Proposition~\ref{P:dimensionality}.
\begin{proof}
Dimensional bounding: if $k>|\mathcal{A}|$ then for any $j\in\{1,\cdots,l\}$ we must have $|\bigcup_{i=0}^{k-1}\{f_j(w_i)\}|\le|\mathcal{A}|<k$ and the claim follows by definition of $d_k$.

Dimensional monotonicity: Fixing $1\le k'\le k$, by Proposition~\ref{P:hypermetric} (symmetry), it suffices to prove the claim for $\iota=\text{\rm id}|_{\{0,\cdots,k'-1\}}$. Namely, we want to show
$$d_k(w_0,w_1,\cdots,w_{k-1})\le d_{k'}(w_0,w_1,\cdots,w_{k'-1}).$$
This however follows immediately from the definition of $d_k$ by observing that for any $j\in\{1,\cdots,l\}$ for which $|\bigcup_{i=0}^{k-1}\{f_j(w_i)\}|=k$ we must in turn have $|\bigcup_{i=0}^{k'-1}\{f_j(w_i)\}|=k'$.
\end{proof}

Proposition~\ref{P:linrec}.
\begin{proof}
It suffices to note that if $w\in W$ is a linear recombinant of $W\setminus\{w\}$ then by definition, for each $j\in\{1,\cdots,l\}$ there exists an $i\in\{0,\cdots,k-1\}$ such that $f_j(w)=f_j(w_i)$. This means that for for each $j\in\{1,\cdots,l\}$ we must have $|\bigcup_{i=0}^{k-1}\{f_j(w_i)\}|\le k-1<k$ and as such $d_k(w_0,\cdots,w_{k-1})=0$ as claimed.
\end{proof}

Proposition~\ref{P:wdef}.
\begin{proof}
This is an immediate consequence of Proposition~\ref{P:dimensionality} (dimensional bounding), for any $k>|\mathcal{A}|$.
\end{proof}

Proposition~\ref{P:wsc}.
\begin{proof}
Again, by Proposition~\ref{P:dimensionality} (dimensional monotonicity), we have $d_{k+1}(\sigma)\leq d_{k'+1}(\tau)$ and so immediately $v_W(\sigma)=\pi^{d_{k+1}(\sigma)}|\pi^{d_{k'+1}(\tau)}=v_W(\tau)$.
\end{proof}

Proposition~\ref{P:shuffle}.
\begin{proof}
The first claim, $(W_\epsilon)^\omega=(W^\omega)_\epsilon$, follows immediately by observing the commutative identity for each entry in the alignment matrix of $W$

$$\begin{tikzcd}
f_j(w_i) \arrow{r}{\omega} \arrow[swap]{d}{\epsilon} & f_{\omega(j)}(w_i) \arrow{d}{\epsilon} \\%
f_j(w_{\epsilon(i)}) \arrow{r}{\omega}& f_{\omega(j)}(w_{\epsilon(i)})
\end{tikzcd}.
$$

For the second claim it suffices to investigate the action a row and column shuffle pair $(\epsilon,\omega)$ has on a fixed $k-1$-simplex $\sigma=\{w_0,\cdots,w_{k-1}\}\in X(W)$.

Note first that $d_k(\epsilon.\sigma)=d_k(\sigma)$ by construction, where $\epsilon.\sigma\in W_\epsilon$ is the simplex in $W_\epsilon$ corresponding to $\sigma$. On the other hand, for $I$ the indicator function in the proof of Proposition~\ref{P:hypermetric}, we have that 
$$d_k(\sigma)=\sum_{j=1}^l I(f_j(w_0),\cdots,f_j(w_{k-1}))=\sum_{j=1}^l I(f_{\omega(j)}(w_0),\cdots,f_{\omega(j)}(w_{k-1}))=d_k(\omega.\sigma).$$

By the previous claim, the order in which we apply the two actions does not matter, and as such $W^\omega_\sigma\ni\sigma^\omega_\epsilon=\omega.\epsilon.\sigma$ with $d_k(\sigma^\omega_\epsilon)=d_k(\sigma)$ and the proposition follows.
\end{proof}

Proposition~\ref{P:dimensionality 2}.
\begin{proof}
This is an immediate consequence of Proposition~\ref{P:wdef} which bounds the dimensionality of the complex.
\end{proof}

Proposition~\ref{P:shuffle 2}.
\begin{proof}
This follows immediately from Proposition~\ref{P:shuffle} which homeomorphically equates the dissimilarity complex of an alignment with its row column shuffle and shows that the arithmetic weight information is preserved under such a transformation.
\end{proof}

Theorem~\ref{T:Vieta}.
\begin{proof}
It suffices to examine the polynomial $P_W(x)=\prod_{i=1}^s (x-b_i)$ where $b_i=|B_i|$ and $s$ is the number of distinct letters that appear in $W$'s column. Vieta's formulae for $P_W(x)$ yield, for each $0\le k\le s$,

$$\sum_{1\le i_0 < i_1 < \cdots < i_k\le s} \left(\prod_{j = 0}^k b_{i_j}\right)=(-1)^k\frac{p_{s-k}}{p_s}
$$

where $p_q$ is the coefficient of the term $x^q$ in $P_W(x)$ for $0\le q\le s$. In this particular case $p_s=1$. If we showed that $p_{s-k}=c_k$ for any $0\le k\le s$ then  $P_W(x)=D_W(x)$ and the theorem would follow. To this end, for $I$ the indicator function in the proof of Proposition~\ref{P:hypermetric}, we can write $$c_k=\sum_{\sigma\in K_k(W)}I(\sigma)=|K_k(W)|.$$

To construct a simplex in $K_k(W)$ it suffices to select $k+1$ bins and select one sequence from each bin. As such, the theorem follows from

$$|K_k(W)|=\sum_{1\le i_0 < i_1 < \cdots < i_k\le s} \left(\prod_{j = 0}^k b_{i_j}\right)=p_{s-k}.$$
\end{proof}

Theorem~\ref{T:bins}.

Note that this Theorem is equivalent to Theorem $4$ in \cite{bolker1976simplicial}. The original proof was based on simplicial joins and a Mayer-Vietoris type sequence. Here we present an alternate, more combinatorial proof (and only for part (b)), consisting of an explicit construction of the relevant generators of the homology at the $s-1$ dimension.

Before we present the closed form formulae of the weighted homology for the single column alignment, let us fix a simplicial ordering. Without the loss of generality, we can pick a simplicial order that is compatible with the bin ordering, see Figure~\ref{F:lattice} (a). Then for any maximal simplex $\sigma=[x_1,\cdots,x_s]$, we have $x_i\in B_i$, and a simplex is now considered an ordered tuple.

\begin{proof}
Since all simplices in $X(W)$ have weight $\pi$, we have $\partial_k^v(\sigma)=\sum_{i=0}^k (-1)^i \frac{\pi}{\pi} \hat{\sigma_i}=\sum_{i=0}^k (-1)^i  \hat{\sigma_i}$.\\

It suffices to find a set of $R$-linearly independent set of generators for $\text{\rm Ker}(\partial^v_{s-1})= H^v_{s-1}(X)$ of size $b$. By construction, $C_{s-1,R}(X)=\langle M\rangle _{R}\neq 0$. Fix $\sigma_0=[x_1,\cdots,x_s]\in M$ and consider\\

\textit{Case 1:} there exists no other simplex $\sigma'=[y_1,\cdots,y_s]\in M$ such that $y_i\ne x_i$ for all $1\le i\le s$. In this case, at least one bin has size $1$ and thus we have $b=0$.

\textit{Claim:} $H^v_{s-1}(X)=0$.

To prove this, consider $c\in \text{\rm Ker}(\partial^v_{s-1})$ and grade it by $|\sigma_0\cap \sigma|$, denoting the number of vertices the two maximal simplices share. In this case, the grading starts at $|\sigma_0\cap \sigma|=1$

$$
c=\sum_{k=1}^s \sum_{|\sigma_0\cap \sigma|=k} a_{\sigma}\sigma.  
$$

Let $\sigma^*\in c$ satisfy $|\sigma_0\cap \sigma^*|=1$. Then $\sigma^*=[y_1,\ldots,x_i,\ldots,y_s]$, for some $1\leq i\leq s$, while $y_j\neq x_j$ for all $j\neq i$. Let $\hat{\sigma^*}_i=[y_1,\ldots,\hat{x}_i,\ldots,y_s]$.
Consider all possible $\sigma^{**}\in M$ with $|\sigma^{**}\cap \sigma_0|\geq 1$, such that $\hat{\sigma^*}_i=\hat{\sigma^{**}}_i\subset \sigma^{**}$. Since $|\sigma^{**}\cap \sigma_0|\geq 1$ and $|\hat{\sigma^*}_i \cap \sigma_0|=0$, we must have $\sigma^{**}=\sigma^*$. Namely, $\sigma^*$ is the only simplex in $M$ that contains $\hat{\sigma^*}_i$ as a face. Then $\partial^v_{s-1}(c)=0\implies a_{\sigma^*}=0$. This holds independently for all $\sigma^*$ with $|\sigma_0\cap \sigma^*|=1$. Therefore we have

$$
c=\sum_{k=2}^s \sum_{|\sigma_0\cap \sigma|=k} a_{\sigma}\sigma.
$$

We proceed similarly for each $k\ge 2$ in order, which eventually leads to $c=0$.\\

\textit{Case 2:} there exist at least one simplex $\sigma'=[y_1,\cdots,y_s]\in M$ such that $y_i\ne x_i$ for all $1\le i\le s$. In this case, each bin must contain at least $2$ vertices, see Figure~\ref{F:lattice} (b). Let
$$L(\sigma'):=\{[z_1,\cdots,z_s]\in M|z_j=x_j\text{ or }z_j=y_j,\text{ for all } 1\le j\le s\},$$ 
with $x_i$ or $y_i$ appearing at the same coordinate since they are chosen from the same bin and the $0$-simplices follow an order that is compatible with the bin order, see Figure~\ref{F:lattice} (c). We make the Ansatz 

$$\beta=\{l(\sigma'):=\sum_{\sigma\in L(\sigma')}(-1)^{|\sigma_0\cap\sigma|}\sigma|\sigma'\in M, |\sigma_0\cap\sigma'|=0\},$$

noting that $|\beta|=\prod_{i=1}^{s}||B_i|-1|=b$.\\ 

\begin{figure}[ht]%
\centering
\includegraphics[width=0.8\textwidth]{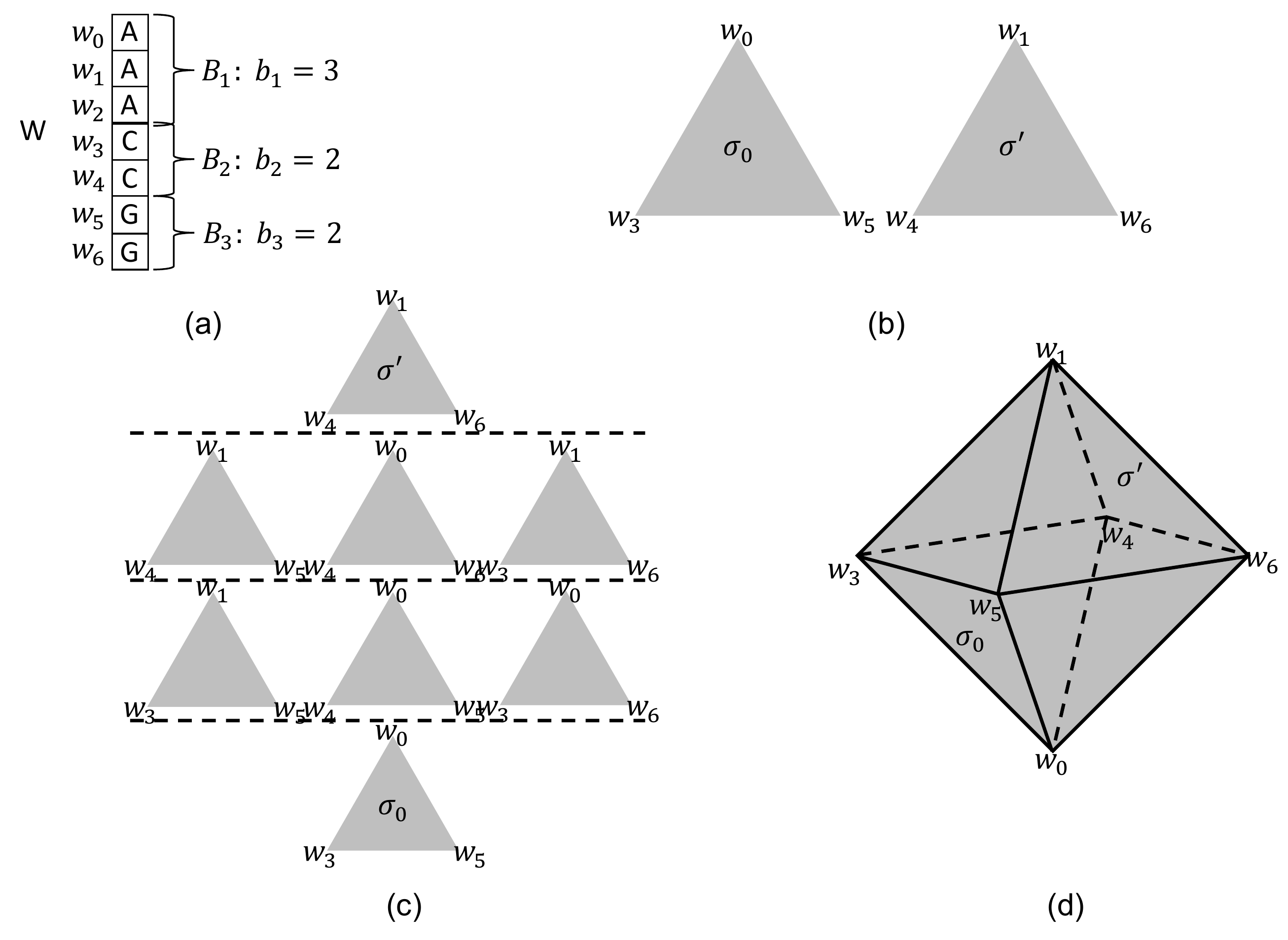}
\caption{(a) A single column alignment $W=[A,A,A,C,C,G,G]$. (b) A fixed $\sigma_0=[w_0,w_3,w_5]$ and $\sigma'=[w_1,w_4,w_6]$, that do not share any vertex in common. (c) All $8$ triangles $\sigma$ in $L(\sigma')$, the grading is given by $|\sigma_0\cap\sigma|$. (d) Geometric illustration of $l(\sigma')$ as an element in $\text{\rm Ker}(\partial^v_{s-1=2})$, the boundary of an octahedron. We have $H^v_{2}(X)=R^{(3-1)(2-1)(2-1)}=R^2$ and the other generator corresponds to $\sigma''=[w_2,w_4,w_6]$.}\label{F:lattice}
\end{figure}

\textit{Claim:} $\beta$ is a $R$-basis for $\text{\rm Ker}(\partial^v_{s-1})$. 

A fixed $\sigma'$ only ever appears in $l(\sigma')$, therefore $\beta$ is a $R$-linearly independent set. 

We next show $\beta\subset \text{\rm Ker}(\partial^v_{s-1})$. Fix $l(\sigma')\in\beta$ and consider

$$
\partial^v_{s-1}(l(\sigma'))=\sum_{\sigma\in L(\sigma')}(-1)^{|\sigma_0\cap\sigma|}\partial^v_{s-1}(\sigma)=\sum_{\sigma\in L(\sigma')}(-1)^{|\sigma_0\cap\sigma|}\sum_i (-1)^i \hat{\sigma}_i.
$$

Note that $\hat{\sigma}_i=[z_1,\ldots,\hat{z}_i,\ldots,z_s]$ appears in exactly two images, $$(-1)^{|\sigma_0\cap\sigma^*|}\partial^v_{s-1}(\sigma^*)\text{ and }(-1)^{|\sigma_0\cap\sigma^{**}|}\partial^v_{s-1}(\sigma^{**}),$$ where $\sigma^*=[z_1,\ldots,x_i,\ldots,z_s]$ and $\sigma^{**}=[z_1,\ldots,y_i,\ldots,z_s]$, and it does so with opposite signs, hence $l(\sigma')\in \text{\rm Ker}(\partial^v_{s-1})$, see Figure~\ref{F:lattice} (d).

Finally, we show $\langle \beta\rangle_{R}=\text{\rm Ker}(\partial^v_{s-1})$. Consider

$$\text{\rm Ker}(\partial^v_{s-1})\ni c=\sum_{\sigma\in K_{s-1}(W)} a_{\sigma} \sigma=\sum_{k=0}^s \sum_{|\sigma_0\cap \sigma|=k} a_{\sigma}\sigma=\sum_{|\sigma_0\cap \sigma'|=0}a'_{\sigma'}\sigma'+\sum_{k=1}^s \sum_{|\sigma_0\cap \sigma|=k} a_{\sigma}\sigma.$$

\textit{Claim: } $c=\sum_{|\sigma_0\cap \sigma'|=0}a'_{\sigma'}l(\sigma')$.

Let
$$\text{\rm Ker}(\partial^v_{s-1})\ni c^*=c-\sum_{|\sigma_0\cap \sigma'|=0}a'_{\sigma'}l(\sigma').$$
By construction, the coefficient of $\sigma'\in c^*$, with $|\sigma'\cap \sigma_0|=0$, is $0$ hence

$$
c^*=\sum_{k=1}^s \sum_{|\sigma_0\cap \sigma|=k} a^*_{\sigma}\sigma.  
$$

Iterating Case $1$, yields $c^*=0$.\\
\end{proof}

\section{Acknowledgements}
We want to thank Thomas Li for comments and discussions.

\bibliography{sn-bibliography}

\end{document}